\documentclass[num-refs]{wiley-article}

\usepackage[utf8]{inputenc} %

\usepackage{graphicx}
\graphicspath{{./figures/}}
\usepackage{varioref}

\usepackage{booktabs}
\usepackage{pdflscape}
\usepackage{mathtools}
\usepackage{amsfonts,amsmath,amssymb,mathrsfs}
\usepackage[space]{grffile}
\usepackage{afterpage}
\paperfield{Systems Engineering}
\fundinginfo{G. Bakirtzis is partially supported by the academic and
research chair \emph{Architecture des Systèmes Complexes} through the following partners: Dassault Aviation, Naval Group, Dassault Systèmes, KNDS France, Agence de l’Innovation de Défense, and Institut Polytechnique de Paris.}

\newtheorem{property}{Property}

\usepackage{natbib}
\newcommand{\Hair}{\ifmmode\mskip1mu\else\kern0.08em\fi}

\begin{document}

\title{Mission Aware Cyber-physical Security}

\author[1]{Georgios Bakirtzis}
\author[2]{Bryan Carter}
\author[3]{Cody H.~Fleming}
\author[4]{Carl R. Elks}
\affil[1]{LTCI, Télécom Paris, Institut~Polytechnique~de Paris}
\affil[2]{University of Virginia}
\affil[3]{Iowa State University}
\affil[4]{Virginia Commonwealth University}
\corraddress{Cody Fleming PhD, Iowa State University, Ames, Iowa, 50011, USA}
\corremail{flemingc@iastate.edu}
\maketitle

\begin{abstract}
Perimeter cybersecurity, while essential, has proven insufficient against sophisticated, coordinated, and cyber-physical attacks. In contrast, mission-centric cybersecurity emphasizes finding evidence of attack impact on mission success, allowing for targeted resource allocation to mitigate vulnerabilities and protect critical assets. \emph{Mission Aware} is a systems-theoretic cybersecurity analysis that identifies components which, if compromised, destabilize the overall mission. It generates evidence by finding potential attack vectors relevant to mission-linked elements and traces this evidence to mission requirements, prioritizing high-impact vulnerabilities relative to mission objectives. \emph{Mission Aware} is an informational tool for system resilience by unifying cybersecurity analysis with core systems engineering goals.

\keywords{mission aware cybersecurity, resilience-security coengineering, cyber-physical systems.}
\end{abstract}%

\section{Introduction}

Cyber-physical systems (CPS) integrate a diverse set of hardware and software components to provide safety-critical service capabilities. These components include computing platforms, sensors, control systems and communication networks to monitor operational conditions and control system assets to complete their underlying mission. A security incident in any of the underlying subsystems can, therefore, destabilize the overall mission of a CPS system and transition the system to a hazardous state. Security analysis throughout the CPS' lifecycle can mitigate hazards and harden the system from mission disturbances. Efforts towards designing and building resilient CPS require security analysis to move towards the beginning of the lifecycle.

The imperative to integrate security considerations early in the system lifecycle is strongly supported by the defense community. Studies indicate that 70-80\% of decisions impacting the efficiency and cost of safety and security are made during the initial concept development phase of any project~\cite{frola_system_1984,saravi_estimating_2008}. Model-based security can give a ``what if'' perspective into a system's security before software is written and hardware has been wired together, better informing designers for the types of choices they ought to make.

Furthermore, while defenders typically rely on list-based security mechanisms, attackers gain a strategic advantage by conceptualizing security in terms of graphs~\cite{lambert:2015}. In particular, perimeter security demonstrates some success in protecting CPS but they tend to be prescriptive (for example, use a firewall, encrypt communication channels, etc.) without being aware of the nature and purpose of the system and its mission---including the possible hazards that might occur in the face of adversity. In contrast, mission-centric cybersecurity presupposes that there is a specific expected service that needs to be protected and, hence, provides adequate justification for defense solutions against attack vectors. Therefore, mission-centric cybersecurity provides awareness of how sophisticated attacks can influence mission success and, consequently, focuses system designers' resources and efforts in mitigating potential vulnerabilities~\cite{jakobson2013mission}.

Based on the above insights, we identify three needs in the area of model-based CPS security: %
\begin{enumerate}
\item deriving strategic resilience from both system and mission contexts; 
\item establishing systematic, risk-based security analysis that transcends traditional compliance checklists and is applicable early in the system lifecycle; and
\item adopting a graph-based representation of security for defenders, aligning their perspective with that of attackers.
\end{enumerate}

\emph{Mission Aware} directly addresses the above challenges by being a proactive, risk-orientated, and graph-theoretic cybersecurity analysis. A strategic approach starts with requirements gathering to understand the mission of the system, what it is expected to do and what potential hazards it might face, who it is expected to serve, and for what purpose. By answering these questions, we derive mapping to systems engineering concepts, such as the mission requirements, the functional behaviors, and the architecture of a system. Mission Aware combines all three domains in a well-formed mission specification based on multiple graphs. Finally, we use public vulnerability repositories (e.g., CAPEC,\Hair\footnote{Common Attack Pattern Enumeration \& Classification, capec.mitre.org} CWE,\Hair\footnote{Common Weakness Enumeration, cwe.mitre.org} CVE,\Hair\footnote{Common Vulnerabilities and Exposures, cve.mitre.org} etc.) to assess the security posture of the critical subsystems that directly relate to potential mission degradation, bridging the gap between security information and systems engineering.  

We have thus far never published a manuscript on the guiding principles and the formalism that led to Mission Aware cybersecurity modeling and assessment. The following work has led to several reports and publications that give hints to this system of thought for cyber-physical security, but have yet to be explicitly exposed as a unit in the systems engineering context. In particular, the principles developed in this manuscript led to a series of work on model-based security.
\begin{itemize}
    \item A series of technical reports on applying Mission Aware to usecases~\citep{horowitz:2017,horowitz:2018,beling:2019,mcdermott:2021,sherburne:2023}.
    \item A series of research papers on model-based security and safety coengineering~\citep{carter:2018,bakirtzis:2018b,bakirtzis:2018a,carter:2019b,carter:2019a,carter:2019c,bakirtzis:2020a,bakirtzis:2020b,bakirtzis:2021d,fleming:2021,beling:2023}.
    \item A series of prototype software tools for model-based security~\citep{georgios_bakirtzis_2018_1308914,georgios_bakirtzis_2020_3766874,georgios_bakirtzis_2018_1318537}.
\end{itemize}

We believe that the exposition of the guiding principles and light formalism behind \emph{Mission Aware} is an important addition to the literature to expose the foundational context of the above work.

\section{Background}

The security community has developed approaches and methods addressing the problem of vulnerability and security assessment varying in intent, scope, and objectives. A research line attempts to exercise concepts from dependable and safe computing to the realm of security. Model-based quantitative security analysis with techniques from dependability are motivated by Nicol et al.~\cite{nicol2014dependability}. Safety and security, however, can often be difficult to assure through a purely quantitative framework~\cite{harkleroad2013risk}, particularly when developing new systems or missions. In contrast, we use STAMP to capture the potential hazardous scenarios the system might face during its deployment and assess the security posture qualitatively by using evidence and the model.

Hong et al.~\cite{hong2017survey} present a  comprehensive survey of graph-theoretic approaches, including but not limited to attack trees, attack defense trees, attack graphs, etc. While our elicitation process and model can assist with any of the above tool-based approaches, we gain new insights by finding the impact to the mission instead of focusing just on the system itself with no awareness of its expected service. Other graph-based approaches target the compliance of policy or standard~\cite{Weaver2013segs}. Further notable work in this area is described by Chen et al.~\cite{Chen2013nspw}, where workflows are used to assure the security of the system in a given scenario. Jauhar et al. produce security argument graphs through CyberSAGE models to assess failure scenarios in the smart grid~\cite{jauhar2015model}.

Systems-theoretic approaches in the area of safety-critical CPS include STPA-Sec~\cite{young_systems_2013} and STPA-SafeSec~\cite{friedberg2016}. The former is a general methodology that can be applied to any critical system. STPA-SafeSec shows the application of STPA-SafeSec in the context of the power grid. In this work we utilize STPA-Sec to identify the hazards and encode them in the model but do not limit ourselves to the method. Instead, in this paper the hazards are captured in a hierarchical model to assess the impact of reported evidence, through vulnerability databases, on a specific system architecture. Additionally, Jones et al. propose a technique called System-Aware that follows a system modeling methodology specific to risk analysis through voting techniques using spreadsheets~\cite{jones2012}.

Ouchani et al. describe a model-based approach to security using SysML~\cite{Ouchani2014pcs}. Their approach relies in representing both the model and the attack in SysML and is simulated within that framework. Lemaire et al. propose a formal verification scheme using vulnerability data and a SysML system description~\cite{Lemaire2017}. Brunner et al. proposed a combined model for safety and security based on Unified Modeling Language (UML) diagrams~\cite{brunner_towards_2017}.

In the context of space missions, Pecharich et al. propose a mission-centric approach to analyzing the security posture of systems and how to use the assessment to make informed decisions about system deployment~\cite{pecharich2016mission}. However, the method and corresponding toolset relies on the construction of attack trees. Similarly, Jakobson presents a more general motivation for a shift from the current methods to mission-centric cybersecurity~\cite{jakobson2013mission,jakobson2014mission}.

Finally, Mead and Woody use existing information about malware to inform, at the early stages of a systems development, the requirements elicitation process~\cite{mead_security}. This constitutes an evidence-based elicitation process but requires knowledge of specific malware architectures and how they are used to infect the system under analysis.

\section{Mission Aware Cybersecurity}

\begin{figure}[!t]
\includegraphics[width=1.0\textwidth]{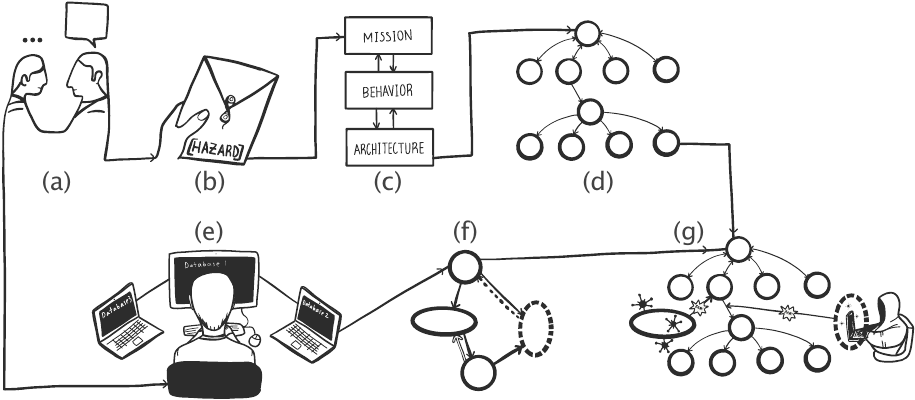}
\caption{The Mission Aware approach conducts requirements elicitation, hazard analysis, SysML modeling, and analyzes the security posture through evidence and graph representation to provide a holistic mission-centric view.}
\label{fig:concepts}
\end{figure}

The Mission Aware methodology, as the name implies, is grounded in understanding the expected service of the system, i.e., the mission, and the goals of the mission from various stakeholders' expert knowledge and operational beliefs (Fig~\ref{fig:concepts}). In this section, we discuss the basic elements and workflow of Mission Aware.

Mission Aware begins with a structured elicitation process from various stakeholders that first defines the mission scenarios and then identifies both the possible mission hazards and the type of threat space that is going to be associated with the system architecture (Fig~\ref{fig:concepts} (a)). The guided stakeholder elicitation is the driving concept in Mission Aware. The typical stakeholders ensemble includes supervisory staff, commanders, operations staff, end users, and subject domain experts.

Following the guided stakeholder elicitation, the methodology branches into two paths. The first path addresses the modeling of the target system architecture with respect to functionality, mission constraints and requirements of CPS, admissible behaviors, restrictions, and hazards of the mission. The second path identifies and assesses threats to the specified mission. Along this path, the Mission Aware methodology discovers relevant attack information at the earliest possible stage of system development. Threat information (e.g., attack patterns, vulnerabilities, and weaknesses) mined from public attack vector databases are used to drive system vulnerability detection and mission impact.

\paragraph{Modeling}
The purpose of the modeling effort in Mission Aware is to capture the stakeholders' beliefs in the form of records and artifacts, provide a systematic framework to analyze these artifacts, and construct a specification in Systems Modeling Language (SysML) that can grow and be modified throughout the lifecycle of the system. Specifically, these artifacts are analyzed in a systematic approach based on Systems-Theoretic Accident Model and Process (STAMP), which is detailed in Section~\ref{sec:stpa}, capturing the potential hazards the mission might face in a tabular format (Fig~\ref{fig:concepts} (b)). The outcomes of this systems-theoretic analysis are then captured into a hierarchical SysML~\cite{hause_sysml_2006} model of the target system. The target system model is segmented into three domains: mission, behavior, and architecture (Fig~\ref{fig:concepts}~(c) and Fig~\ref{fig:hierarchy}). The SysML model can be further extended and modified to represent changes in the mission of the system, the behaviors, as well as architectural changes.

\paragraph{Attack Vector Space}
The attack vector space allows us to provide real information about attack patterns, weaknesses, and vulnerabilities associated with the model. In particular, the attack vector space find the hazards that might cause mission degradation but also assess how likely the threat is based on evidence. Using the elicited information, the analyst is also informed on which databases are applicable to the system and its corresponding mission (Fig~\ref{fig:concepts} (e)). Through the databases, we extract vulnerability information that can potentially violate the mission-level requirements and the corresponding system architecture (Fig~\ref{fig:concepts} (f)). This step can be complemented with further information after the system architecture has been modeled.

\paragraph{Impact of attacks to mission success}
The final step in Mission Aware is to ascertain through the model and its corresponding attack vector space if there can be significant impact on the mission. To assess this impact and relate attacks to potential hazards, the outputs of both paths are transformed into graph metamodels that represent the mission specification and its associated evidence. The use of graph metamodels is important for two specific reasons. First, graphs in Mission Aware carry important model attribute information that can map to attack vectors of the components of a system. Second, graphs are a common formalism for both attack and system models and have been proven practical in several contexts in cybersecurity~\cite{hong2017survey}. This allows us to assess the impact of an adversarial event by finding the paths that violate mission-level requirements (Fig~\ref{fig:concepts} (g)).

The full analysis can be repeated as many times as needed to provide confidence in mission success by feeding back the results (mission impact) to the stakeholders.
\subsection{Requirements \& Mission Information Elicitation}

\begin{figure}[!t]
\centering
\includegraphics[width=.7\linewidth]{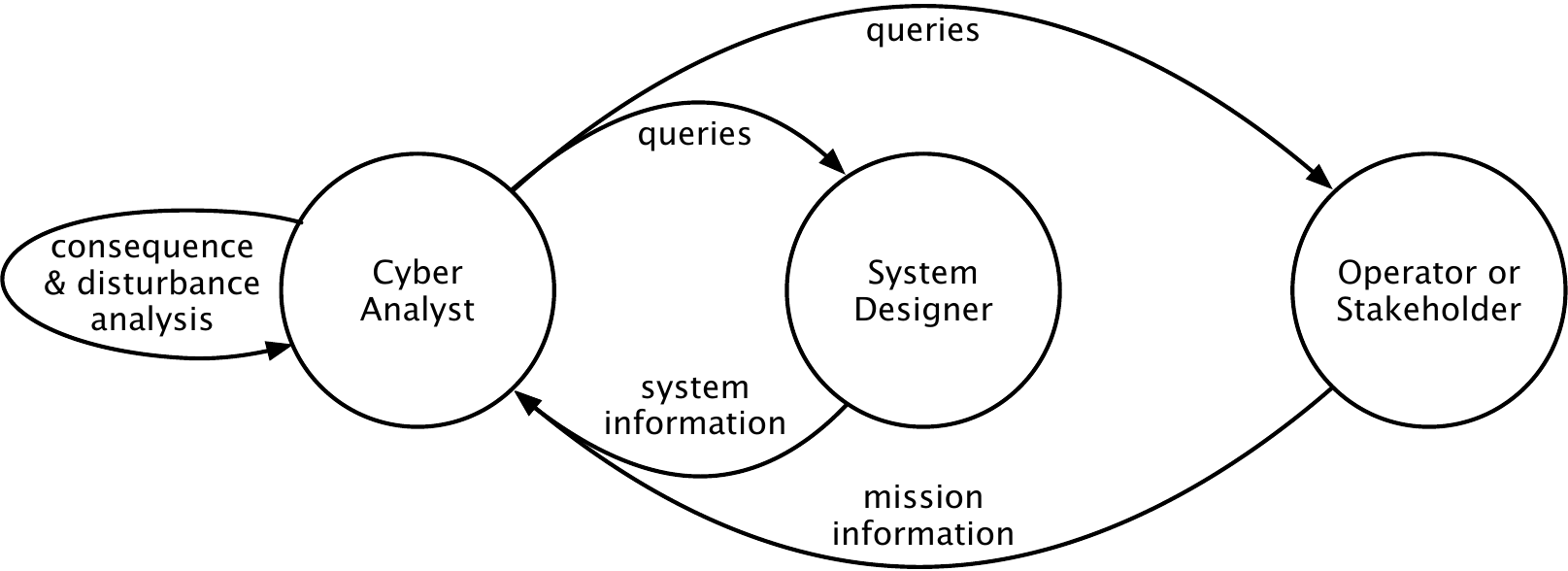}
\caption{The guided stakeholder elicitation seeks different information from varying stakeholders.}
\label{fig:war-room}
\end{figure}

The guided stakeholder elicitation activity leverages the strengths of different stakeholders to identify mission goals, objectives, unacceptable outcomes, expectations, and procedures. It also makes an initial prognosis for the threat level that the mission and the system might combat. Such stakeholders include the mission owner, the system operator, attack analysts, and the acquisition cost experts. Analysts engage these stakeholders by asking a series of queries and proposing hypothetical scenarios to generate the information listed above (figure~\ref{fig:war-room}). For example, an analyst may ask a UAV operator how they might handle the malfunction of an imaging payload, or ask an attack analyst how mission timing can affect the types of threats to the system. Such questions serve as a starting point to define the mission's operational boundaries and unacceptable outcomes. This information is critical for establishing a baseline for the subsequent hazard analysis, rather than limiting the analysis to only those failures the operator has previously considered. By maturing from the \emph{fuzzy front end} of a narrative description to a rough block diagram to a robust system model, we are able to concretely capture the diverse information in a single SysML model.

Through the guided stakeholder elicitation exercise, we are able to encode mission requirements as high-level properties to be preserved. Additionally, through expert input about mission objectives and unacceptable outcomes we are able to pinpoint and classify critical components of a system, their interaction with users and the environment, and the relationship between those critical assets and the mission. %
The stakeholder elicitation directly informs the subsequent hazard analysis in the next step of the Mission Aware process with unacceptable outcomes, potential hazards, and procedural behaviors. This is advantageous as this information is now produced explicitly by the stakeholders rather than derived or assumed by the analysts. The results of the guided stakeholder elicitation exercise are also important in deciding which databases are applicable to the mission and its corresponding system based on the threat analysis information provided by the stakeholders. %

\subsection{Disturbance and Consequence Modeling}
\label{sec:stpa}

Systems theory views a system as a hierarchy, with each level imposing a set of constraints on the behavior of the level below. When these constraints break down, so does the safety and security of the system. STAMP~\cite{leveson_engineering_2012} is an accident causality model that describes these breakdowns. In the context of Mission Aware, an adversary exploits vulnerabilities to ultimately change the behavior of a system and affect the outcome of a mission. STAMP allows us to analyze unintended system behavior as a control problem rather than a component failure problem, which enables a holistic investigation of both safety and security. 

More specifically, STAMP extends traditional causality modeling beyond component failures to include human interactions, the environment, organizational structures, and hazardous interactions with non-failing components. Inadequate control or handling of these factors lead to losses, not just simple component failure. This concept proves useful for the Mission Aware approach as it shifts focus from preventing failures to enforcing constraints; in other words, it helps strategize how functionality can be preserved in the face of disruptions, rather than attempting to prevent all disruptions. 

While STAMP mainly focuses on system safety, its principles can be applied to help foster a holistic approach to increasing system security and resilience. Although both safety and security analyses focus on preventing system-level losses, their distinction lies in the assumed intent of the disturbance. However, STPA-Sec is an extension of STAMP that applies the same concepts of enforcing safety constraints to control vulnerabilities~\cite{young_systems_2013,carter:2018}. 
STPA-Sec shifts the focus of STAMP from designing against unintentional disruptions (safety) to protecting against intentional disruptions (security). More specifically, STPA-Sec operates from the top-down by outlining unacceptable losses or outcomes within the mission to establish clear priorities that both guide later analysis as well as influence potential future solutions. Furthermore, the STPA-Sec model systematically encodes these unacceptable losses, the hazardous conditions that could lead to these losses, and the control actions and circumstances under which these actions become hazardous. 

The STPA-Sec model generates the links between the mission requirements and information from stakeholder elicitation and the behavior of the system while attempting to complete that mission. The stakeholder inputs about the mission directly feed into the top levels of the STPA-Sec analysis, and in particular the unacceptable losses and the prioritization of each. Furthermore, the behavior and functionality described in the model provides the basis for a model of the system's architecture which helps create the traceability between hardware and software through mission requirements. This allows the analysis to identify and evaluate potential vulnerabilities with respect to their effects on the outcome of the mission rather than blindly trying to eliminate gaps in security, a key tenet to the Mission Aware approach.

\subsection{Hierarchical Systems Modeling}

\begin{figure}[!t]
\centering
\includegraphics[width=0.4\linewidth]{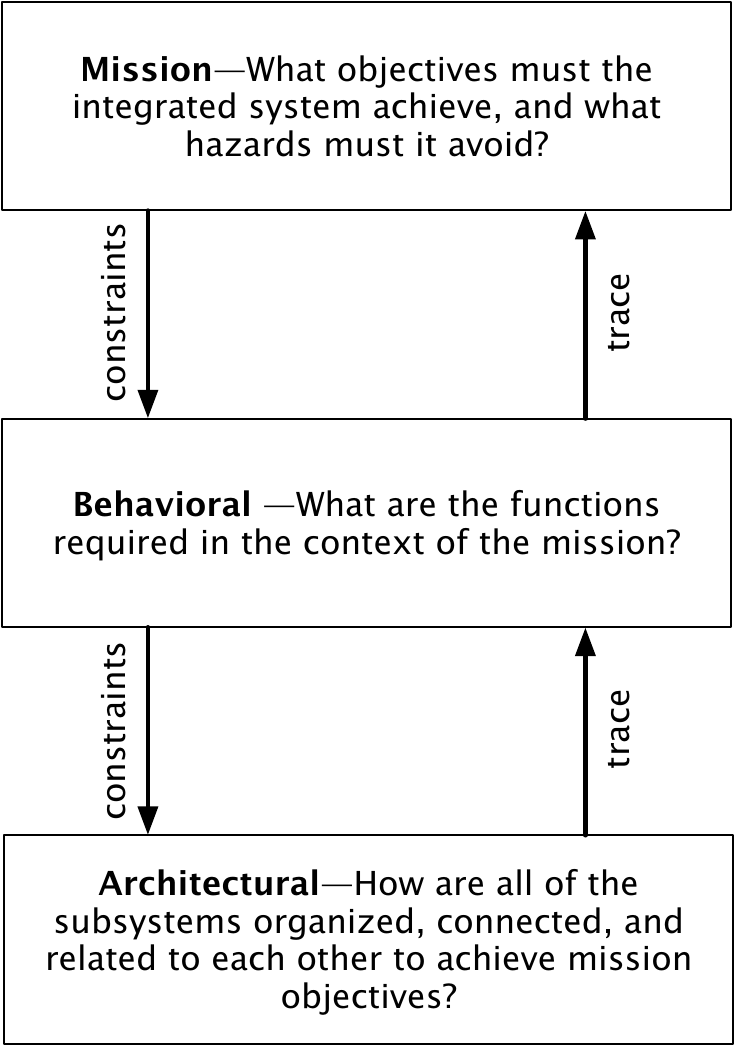}
\caption{The hierarchy is modeled top-to-bottom and traced bottom-to-top.}
\label{fig:hierarchy}
\end{figure}

System models are developed to simultaneously achieve two objectives. The first is to ensure fidelity to an actual system's (or system-of-systems') behavior. The second is to allow the model to be ``virtually attacked,'' and it is this latter objective that requires additional semantics compared to a typical SysML modeling effort.

The SysML modeling starts by constructing the mission requirements, system functions, and system structure. These are the baseline models that correspond to each of the domains (figure~\ref{fig:hierarchy}) and include all the attributes and behaviors above. Mission requirements, system functions, and system structure are as encoded in the following standard SysML diagram types, respectively: requirements diagrams; activity diagrams, which also encode behavior related to STAMP and STPA from Sec~\ref{sec:stpa}; and block definition and internal block diagrams.

A second modeling pass completes the mission specification by tracing mission requirements to system functions and system structure (in SysML this is achieved by extending the requirements diagrams). This allows an analyst to identify vulnerabilities of system components and then find all trace paths that impact mission requirements and better understand the security needs not only based on the utilized system but also on the mission requirements. Ideally it causes a reduction of the architectural elements that need to be considered in the security evaluation; however, in the worst-case scenario a mission might be impacted by all elements in the system architecture model.

In order for the model to be ``attackable'' we construct the following semantics, i.e., descriptors and behaviors, that capture:

\begin{itemize}
\item \textbf{Information flow:} Types of information flow for
  \textit{Input} and \textit{Output} ports.
\item \textbf{Properties:} Information flows or components associated with
  confidentiality, integrity, availability, and also, restrictions,
  sharing, etc.
\item \textbf{Functionality:} What function a sub-system has relative
  to the whole system and what service it provides
\item \textbf{Non-functional attributes:} Timeliness, responsiveness, user
  interaction, etc.
\item \textbf{Interface interactions} How a component interacts with users
  and entities.
\end{itemize}

These semantics allow us to map potential attack vectors using the architectural specification of the system. This in turn, provides evidence that the impact to the mission specification is realizable based on historic reported attack patterns, weaknesses, and vulnerabilities.

\section{Mission Aware Graph Metamodels}
\label{sec:graphs}

In this section, we construct a light graph-theoretic formalism that act as metamodels for assessing the overall security posture of CPS based on their mission-level requirements, admissible behaviors, and system architecture. The adoption of graph metamodels in Mission Aware provides a standardized framework, enabling its application across diverse toolsets and enhancing interoperability in complex systems analysis. 

\subsection{Model}

We produce a common language formalism for security assessment for mission-centric CPS. This formalism  makes connections to the SysML encapsulation and the hierarchy produced by the systems-theoretic approach. It also introduces the notions of attack chain and mission impact. The following terminology is one of potentially several ways to represent the concepts present in this paper. We make explicit the types along the wires by augmenting the usual graph presentation with source and target types. These help us disentangle the relationships between different elements in the hierarchy of the model; that is, requirements, functions, structure.

\begin{definition}[Mission Requirements]
The mission requirements of a cyber-physical system is a graph \(R \coloneqq (V_R, A_R, \textit{src}_R, \textit{tgt}_R)\), where \(V_R\) the vertices of \(R\); \(A_R\) the arrows of \(R\); \(\textit{src}_R, \textit{tgt}_R: A_R \rightarrow V_R\). \(V_R\) represents the requirements; \(A_R\) the relations of each requirement; and \(\textit{src}_R, \textit{tgt}_R\) the type of relation, i.e., prerequisite or refinement.
\end{definition}

\begin{definition}[System Function]
The function of a cyber-physical system is a graph \(F \coloneqq (V_F, A_F, \textit{src}_F, \textit{tgt}_F)\), where \(V_F\) the set of verices of \(F\); \(A_F\) the set of arrows of \(F\); and \(\textit{src}_F, \textit{tgt}_F: A_F \rightarrow V_F\). \(V_F\) represents the admissible behaviors, i.e., functionality, non-functional attributes, and interface interactions, based on the requirements in \(R\), \(A_F\) defines the admissible decision flow allowed by \(R\), and \(\textit{src}_F, \textit{tgt}_F\) the possible directionality between admissible decisions that exist in \(F\).
\end{definition}

\begin{definition}[System Structure]
\label{def:system}
The structure of a cyber-physical system is a graph \(\Sigma \coloneqq (V_\Sigma, A_\Sigma, \textit{src}_\Sigma, \textit{tgt}_\Sigma, \mathcal{D}_\Sigma)\), where \(V_\Sigma\) is the set of vertices of \(\Sigma\); \(A\) is the set of arrows of \(\Sigma\); \(\textit{src}_\Sigma, \textit{tgt}_\Sigma: A_\Sigma \rightarrow V_\Sigma\), and \(\mathcal{D}_\Sigma\) is the set of descriptors of \(\Sigma\). \(V_\Sigma\) represents the components of a cyber-physical system that implement the admissible behaviors defined by \(F\), \(A_\Sigma\) the information flow communication links to fully implement the admissible behaviors defined in \(F\), \(\textit{src}_\Sigma, \textit{tgt}_\Sigma\) the directionality of the possible cyber or physical interactions between components, and \(\mathcal{D}_\Sigma\) the associated cyber attributes, i.e., properties, for a given vertex or interaction that map to potential attack vectors. %
\end{definition}

In this instance, the mission requirements are encoded in requirement diagrams, the system function is encoded in activity diagrams, and the system structure of the CPS is encoded in internal block and block definition diagrams. Thus, we make use of all three fundamental categories in SysML, namely, behavior, requirements, and structure diagrams. These define the \textit{primitive artifacts} in Mission Aware.

\begin{definition}[Mission Specification]
\label{def:specification}
The mission specification of a cyber-physical system is a graph \(S \coloneqq (V_S, A_S, \textit{src}_S, \textit{tgt}_S, \mathcal{D}_S)\), where \(V_S \subseteq V\) such that \(V = V_R \cup V_F \cup V_\Sigma\) completing the specification by including all the information presented in the mission requirements, how those relate to the system function, and finally how they are realized through the system structure; \(A_S\) the set of arrows that define associativity between any of the elements presented in the vertices that compose the mission, i.e., mission requirements, system function, and system structure; \(\textit{src}_S, \textit{tgt}_S: A_S \rightarrow V_S\) define the directionality of the arrows in \(S\), and \(\mathcal{D}_S\) the descriptors derived from system structure, \(\Sigma\), such that \(\mathcal{D}_S \subseteq \mathcal{D}_\Sigma\).
\end{definition}

The implication of the mission specification definition is that once all \textit{primitive artifacts} are constructed, the mission requirements are traced to their necessary behavioral and architectural elements. This constructs a fully traceable model without changing the completeness of the model. This way, we are able to potentially reduce the number of architectural components that are assessed for their security posture. It follows that if the analysis is systematic, the number of security tools or resilient preemption and mitigation strategies can be reduced to just the architectural components that are present in \(S\) and not necessarily the full graph \(\Sigma\).

\begin{definition}[Attack Vector Space]
\label{def:space}
An attack vector space is a graph \(\textit{AV} \coloneqq (V_{\textit{AV}}, R_{\textit{AV}}, \textit{src}_{\textit{AV}}, \textit{tgt}_{\textit{AV}}, \mathcal{T}_{\textit{AV}})\), where \(V\) the set of vertices of \(\textit{AV}\); \(R_{\textit{AV}}\) is the set of relationships of \(\textit{AV}\), \(\textit{src}_{\textit{AV}}, \textit{tgt}_{\textit{AV}}: R_{\textit{AV}} \rightarrow V_{\textit{AV}}\), in which \(V_{\textit{AV}}\) represents the attack vectors, \(R_{\textit{AV}}\) the related attack vectors of a given component, \(\textit{src}_{\textit{AV}}, \textit{tgt}_{\textit{AV}}\) the directionality of the relationship, that is which one is more abstract, and \(\mathcal{T}_{\textit{AV}}\) the possible types of a relationship between two vertices.
\end{definition}

We model possible attacker actions through the attack vector graph defined above. This model captures all information relevant to the system (as defined by the stakeholders) and its mission but is largely a superset that is constructed by using public vulnerability repositories. For example, in the case that CAPEC, CWE, CVE are used for the analysis, then the vertices, \(V_\textit{AV}\), will be instances of each of the databases and the arrows \(R_\textit{AV}\) the relations between entries. We introduce the concept of types, \(\mathcal{T}_\textit{AV}\), that further categorize what a relation means. These can take the form of intrarelationship if they are within the same database and interrelationship if they are not. We assume that the text that describes each entry is encoded in each vertex within \(V_\textit{AV}\).

\subsection{Finding Applicable Attacks}

One of the major subgoals of Mission Aware is to assess the security posture of a CPS model so that we can ultimately produce systems that are secure by design with respect to mission objectives. To do so we need to be able to map attack vectors from the attack vector space to subsystems from the system structure. Therefore, we require a clear formalism that defines the path in which attacks match subsystems. We term such a path the attack chain, which not only defines orphan nodes or single edges but also, possible sequential attacks that could lead to full mission degradation even if the attacks applied singularly would not.

\begin{definition}[Evidence]
The evidence associated with a given system structure vertex or arrow is a function \(\textit{evidence}: \mathcal{D}_S \rightarrow \mathscr{P}(V_{\textit{AV}})\), where \(\mathcal{D}_S\) is the set of descriptors specified in \(S\) and \(\mathscr{P}(V_{\textit{AV}})\) the power set of the vertices describing the attack vectors \(V_{\textit{AV}}\) in \(AV\).
\end{definition}

An immediate consequence of the definition for evidence is that the set of attacks associated with a descriptor, \(\mathcal{D}_S\), will produce a variety of evidence, some of which is relevant and some of which is irrelevant. Specifically, some evidence is applicable to the subcomponent and therefore the system architecture, while some of it will be false-positive entries associated with that system, \(\Sigma\), and specifically with some descriptor, \(\mathcal{D}_S\).

\begin{definition}[Relevant Evidence] 
Relevant evidence is evidence that is truly applicable to the mission and its corresponding system, i.e. the true-positives that result from the function \(evidence\). A piece of evidence's relevance is a function of both the descriptions in the attack vectors and the descriptions of the system components. However, an attack, \(V_{\textit{AV}}\), such that \(D_S\) is in \(\mathscr{P}(V_{\textit{AV}})\), may not be applicable due to other aspects of the system component that contains \(D_S\). Therefore, relevant evidence is a function \(\textit{rel-evidence}: \mathcal{D}_S \times V_{\textit{AV}} \times V_{\Sigma} \rightarrow E\), where \(E \subseteq \mathscr{P}(V_{\textit{AV}})\).
\end{definition}

Filtering relevant evidence from the larger set of all potential evidence is crucial for the tractability of \emph{Mission Aware}. This is not an arbitrary set of choices but is guided by a set of heuristics. In particular, these include the following:
\begin{enumerate}
\item Architectural context, where a vulnerability is filtered if it does not apply to the specific hardware, operating system, or software libraries present in the system model.
\item Operational context, which discounts threats requiring conditions (e.g., prolonged physical access) that are inconsistent with the mission's deployment scenario.
\item Temporal context, which filters deprecated vulnerabilities for software versions not in use. 
\item Stakeholder input, aligning the threat analysis with the anticipated threat space defined during the initial elicitation. A piece of evidence's relevance is a function of both the descriptions in the attack vectors and the descriptions of the system components.
\end{enumerate}

A descriptor, \(D_S\), is unique and applies to a single element of a system; even if multiple components have the same or similar attributes, a namespace demarcates the descriptor for those components.%

\begin{property}[Attack Chain]
Let the system structure graph be \(\Sigma \coloneqq (V_\Sigma, A_\Sigma, \textit{src}_\Sigma, \textit{tgt}_\Sigma, \mathcal{D}_\Sigma)\). An attack chain \(n\) in \(\Sigma\) is a path, denoted \(p \in \text{Vulnerable Path}_{\Sigma}^{(n)}\), which is a head-to-tail sequence 
\[p = (v_0 \xrightarrow{a_1} v_1 \xrightarrow{a_2} v_2 \xrightarrow{a_3} \cdots \xrightarrow{a_n} v_n)\]

\noindent
of arrows in \(\Sigma\), denoted \({}_{v_0}\left[a_1, a_2, \ldots, a_n\right]\). This path is constructed if and only if there exists an attack vector belonging in the attack vector space that can attach to a given arrow or vertex. Any given set of vulnerable paths in \(\Sigma\) is denoted by:

\[\text{Vulnerable Path}_{\Sigma} \coloneqq \underset{n \in \mathbb{N}}\bigsqcup \text{Vulnerable Path}_\Sigma^{(n)}\]

A single attack at any given vertex defines a trivial path \({}_v\left[ \, \right]\), while a single attack on an edge defines a path of length \(1\). These define the canonical isomorphisms \(\text{Vulnerable Path}_{\Sigma}^{(0)} \cong V_\Sigma\) and \(\text{Vulnerable Path}_{\Sigma}^{(1)} \cong A_\Sigma\).
\end{property}

\subsection{Mission Impact}

An important notion in Mission Aware is that of mission impact. This impact trace spans across the \textit{primitive artifacts} that define the mission specification and can inform analysts and stakeholders about the possible mission-level violations in the presence of a specific attack or attack combination. This is a property of the top-to-bottom modeling approach, which is encoded in the mission specification graph \(S\).

\begin{property}[Impact Trace]
Let the mission specification graph be \(S \coloneqq (V_S, A_S, \textit{src}_S, \textit{tgt}_S, \mathcal{D}_S)\). Then, the impact trace \(m\) is a path in \(S\) that spans across the architecture, functions, and requirements associated with the mission specification.  This path is denoted \(i \in \text{Impact Trace}_{S}^{(m)}\), which is a head-to-tail sequence

\[i = (v_0 \xrightarrow{a_1} v_1 \xrightarrow{a_2} v_2 \xrightarrow{a_3} \cdots \xrightarrow{a_n} v_n)\]

\noindent
of arrows in \(S\), denoted \({}_{v_0}\left[a_1, a_2, \ldots, a_n\right]\). This path is constructed if and only if there exists evidence for a given architectural element in \(V_S \subseteq V_\Sigma\). Any given set of impact traces in \(S\) is denoted by:

\[\text{Impact Trace}_{S} \coloneqq \underset{m \in \mathbb{N}}\bigsqcup \text{Impact Trace}_S^{(m)}\]
\end{property}

Ultimately, the set of impact traces, \(\text{Impact Trace}_{S}\), is the result of the analysis that is reported to the stakeholders and facilitates which preemption and mitigation strategies will be utilized to assure mission success.

\section{Evaluation}
We evaluate the methodology in a military use case, specifically an Unmanned Aerial Vehicle (UAV) reconnaissance mission. We start by producing the results of the stakeholder elicitation structured discussion and continue by applying STPA-Sec hazard analysis to identify the losses, hazards, and control actions associated with the mission. This allows us to create representative models of the mission, behavior, and structure in requirements diagrams, activity diagrams, and internal block definition and block definition diagrams in SysML respectively, which we then represent as graphs through GraphML. Through the graph structures we produce the evidence for the mission-critical components, potential vulnerable paths, and impact traces.
\afterpage{
\begin{landscape}
\begin{table*}
\renewcommand{\arraystretch}{1.3}
\caption{A fragment of unacceptable losses for a UAV reconnaissance mission produced by STPA-Sec}
\label{tab:losses}
\centering
\begin{tabular}{@{}ll@{}}
\toprule
Unacceptable Loss& Description\\
\midrule
L1 & Loss of resources, e.g., human, matériel, due to inaccurate, wrong, or absent information\\
L2& Loss of classified or otherwise sensitive technology, knowledge, or system(s)\\
L3 & Loss of strategically valuable matériel, personnel, or civilians due to loss of control of system(s)\\
\bottomrule
\end{tabular}
\end{table*}

\begin{table*}
\renewcommand{\arraystretch}{1.3}
\caption{A fragment of hazards that can cause unacceptable losses produced by STPA-Sec}
\label{tab:hazards}
\centering
\begin{tabular}{@{}l p{5cm} p{5cm}@{}}
\toprule
Hazard& Worst-case Environment& Associated Losses\\
\midrule
H1---Absence of information& Imminent threat goes undetected& L1: Manpower, matériel, territory, etc.\\ 
H2---Distributing wrong or inaccurate information& Threat is incorrectly identified\par or characterized& L1: Manpower, matériel, territory, etc.\\
H3---Loss of control in unacceptable area& UAV is lost in enemy territory and suffers minimal damage in crash/landing& L2, L3: Compromise of critical systems, intelligence, and/or other potentially classified information or technology\\
\bottomrule
\end{tabular}
\end{table*}

\begin{table*}
\renewcommand{\arraystretch}{1.3}
\caption{A fragment of hazardous control actions at the component level produced by STPA-Sec}
\label{tab:actions}
\centering
\begin{tabular}{@{}p{3cm} p{3.5cm} p{3.5cm} p{4cm} p{3.5cm}@{}}
\toprule
Control Action & Not Providing\par Causes Hazard & Providing\par Causes Hazard & Incorrect Timing\par or Order & Stopped Too Early\par or Applied Too Long \\
\midrule
CA 4.1\par Move control surface & H1, H2, H3: UAV does not avoid inappropriate area,\par or field of view not\par adjusted properly & H1, H2, H3: UAV enters inappropriate area 
  & H1, H2, H3: UAV fails to avoid inappropriate area & H1, H2, H3: UAV temporarily enters inappropriate area \\
CA 4.2\par Take picture\par or collect data & H1, H2: Needed information not collected  & H1, H2: Wrong information collected & H1, H2: Needed information not collected & H1, H2: Needed information not collected or inadequate information collected \\
CA 4.3\par Send data/feedback & H1, H2: Information not supplied to controller & H2, H3: Wrong information sent to controller & H1, H2, H3: Information not sent to controller at correct time & H1, H2, H3: Inadequate information sent to controller \\
\bottomrule
\end{tabular}
\end{table*}

\begin{table*}
\renewcommand{\arraystretch}{1.3}
\caption{Safety Constraints for a fragment of component-level control actions.}
\label{tab:safety}
\centering
\begin{tabular}{@{}p{4.5cm} p{8cm}@{}}
\toprule
Related Control Action & Safety Constraint\\
\midrule
SC 4.1\par Move Control Surface & Control surfaces shall only move upon receiving\par authentic commands from the flight control system\\
SC 4.2\par Take Picture or Collect Data & Data collection shall only occur upon authentic command\par from the operator\\
SC 4.3\par Send Data/Feedback & The component shall relay collected data or send feedback\par to the appropriate monitors at regular intervals\\
\bottomrule
\end{tabular}
\end{table*}
\end{landscape}
}

\subsection{Experimental Setting}

We start by producing the stakeholder elictation artifacts through a structured elicitation of information. First, we produce the mission and system description by conducting a structured discussion between stakeholders. (In this case segmented to groups acting as military commanders, system designers, and analysts.) Once this is completed the analysts further query the rest of the mission stakeholders about the specific mission objectives, potential causes of attack vector impact, and other various hypothetical scenarios, such as impact of system functionality loss. Through this structured discussion the stakeholders provide the analysts with the possible threat-space associated with the specific mission. This recorded, informal information produced by the guided stakeholder elicitation activity allows the analysts to produce a systems-theoretic hazards analysis where a more formal model begins to emerge to describe potential consequences and disturbances to the mission given a system fault (intrinsic to the system or constructed maliciously by an intelligent threat actor).

\begin{figure*}[!t]
\includegraphics[width=1.0\textwidth]{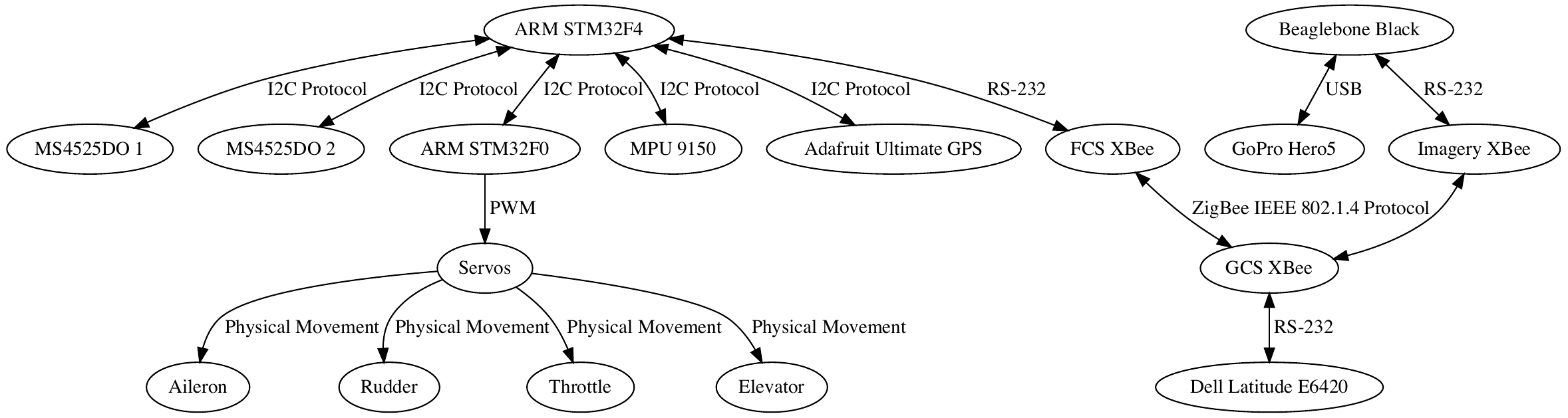}
\caption{The \(\Sigma\) graph represents the architectural topology of the system. This graph is initially constructed in SysML in internal block and block definition diagrams and then extracted to graph form for analysis. (Further information, e.g., descriptors, are not visualized but are accessible through the schema.)}
\label{fig:Sigma}
\end{figure*}

The output of the hazard analysis is then represented in a tabular format. The information present in those tables is encoded in the SysML model of the mission requirements in requirements diagrams, admissible functional behaviors in activity diagrams, and system architecture in block definition diagrams and internal block diagrams using the standard semantics provided in each diagram definition. The subsequent modeling step is to produce the complete mission specification by creating the traces between all model domains in the requirements diagram to encode the rest of the information provided by the stakeholders and hazard analysis.

Finally, following the definitions in Section~\ref{sec:graphs}, we represent the mission requirements as the graph \(R\), the system functions as the graph \(F\), and the system structure as the the graph \(\Sigma\). Through these primitive artifacts we construct the mission specification as the graph \(S\) and use the system structure \(\Sigma\) to find all evidence that can violate the mission objectives and admissible behaviors. This evidence, constructed through cve-search,\Hair\footnote{CVE-SEARCH PROJECT, cve-search.org} is then assessed as relevant or irrelevant by the analysts, the relevant evidence then constructs the attack vector space graph, \(AV\). Relevant evidence from \(AV\) is then used to find potential vulnerable paths, \(\text{Vulnerable Path}_\Sigma\) in the system structure, \(\Sigma\) that can violate mission objectives in \(S\) and produce the impact traces with the highest likelihood of mission degradation, \(\text{Impact Trace}_S\). 

\subsection{Example Analysis}

The guided stakeholder elicitation exercise begins with the definition of a tactical reconnaissance mission utilizing a small UAV with an on-board imagery payload. This particular mission requires that military commanders receive visual information about enemy activities, or lack thereof, in an area of interest to support other future or ongoing missions. Mission failures results from the absence of the reconnaissance information that the UAV would provide. Thus, the GPS and imaging payload are the critical components to mission success since the visual information needs to be linked to a location. Furthermore, vehicle loss or capture is of little concern if the reconnaissance information is relayed up to that point because of the UAV's small size and lack of strategically sensitive information. The military commanders indicate that failing to receive the reconnaissance information would require either a manned secondary mission or the supported mission go without the information; both are undesirable alternatives. 

The most likely threats are a function of inputs from the stakeholders and the experience of the attack analysts. In this case, denial-of-service style attacks pose the greatest threat to the failure of the UAV mission and would likely be an adversary's preferred method of attack. This is due to the supportive nature of the tactical reconnaissance mission, which would give an adversary little time to be aware of the mission or plan a persistent attack to meaningfully impede mission success.  

Using the information from the guided stakeholder elicitation, we conduct consequence and disturbance analysis using STPA-Sec as described in Section~\ref{sec:stpa}. In this mission, the information collected and distributed by the UAV is of more importance than the UAV itself, hence the top priority unacceptable loss is the loss of life or other resources due to the lack of or inaccuracy of the UAV reconnaissance as indicated by the  stakeholders. Furthermore, the stakeholders informs the remaining unacceptable losses, which are placed in order of decreasing priority (Table~\ref{tab:losses}). Next, the STPA-Sec analysis produces a set of hazardous conditions that could lead to these unacceptable losses (Table~\ref{tab:hazards}). These hazards do not necessarily lead to the unacceptable losses mentioned above; the occurrence of these hazards, however, is an indicator of possible full mission degradation. 

After identifying the hazards that could lead to an unacceptable loss, we identify the conditions under which a particular control action becomes hazardous. These control actions show how different actors or controllers within the system alter the behavior of a lower level component or controlled process. For example, the UAV pilot sets and controls the flight plan of the UAV. For the purposes of this paper, a fragment of the control actions and their hazardous circumstances are outlined (Table~\ref{tab:actions}). The component-level in this case refers to the UAV's subsystems: its control surfaces and payload. 

\begin{figure*}[!t]
\includegraphics[width=1.0\textwidth]{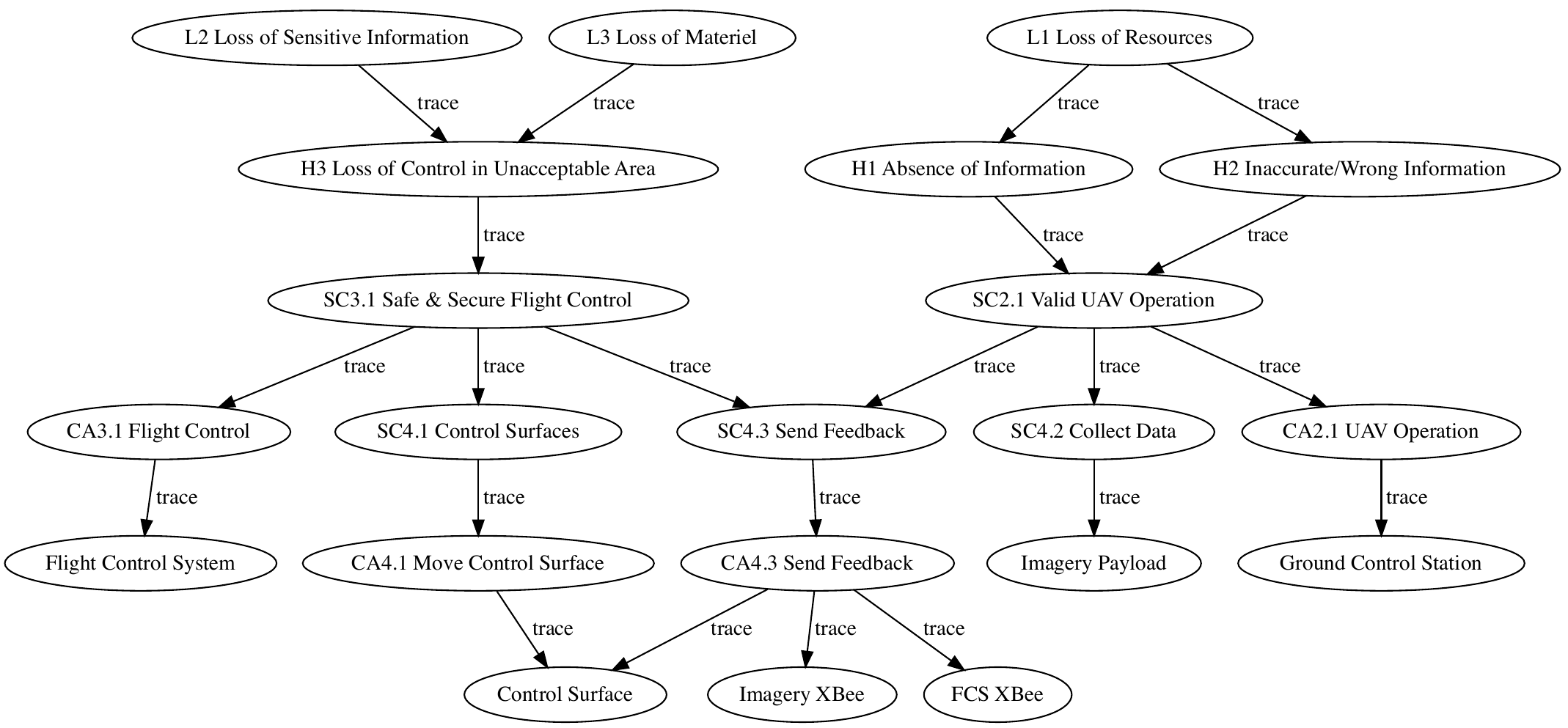}
\caption{The \(S\) graph represents the full mission specification including the requirements, the subset of applicable admissible behaviors, and the subset of subsystems providing these behaviors. This graph is a one-to-one mapping with the SysML model but some of the information, e.g., mission requirement text or structure descriptors \(\mathcal{D}_\Sigma\) are encoded as attributes that can be accessed but are not visually shown. The losses (L) and hazards (H) are encoded in the requirements diagram and are derived from \(R\). The safety constraints (SC) and control actions (CA) are encoded in activity diagrams and are derived from \(F\). All other elements are part of the system structure \(\Sigma\), which is encoded in block definition and internal block definition diagrams. Trace interactions are top-to-bottom following our modeling methodology, while impact is measured bottom-to-top.}
\label{fig:S}
\end{figure*}

We further identify a set of safety constraints for the subset of control actions (Table~\ref{tab:safety}). These constraints serve as a mechanism to help prevent the system from entering a hazardous state without the consent of the operator. This information is modeled in SysML\Hair\footnote{The full SysML model will be distributed online and referenced here. Because of the size of the diagrams we cannot include them here.} as system behavior that needs to be preserved in order to ensure mission success. While SysML provides a common framework to construct and modify models by Systems Engineers, our analysis is exercised by utilizing the graph structures and the evidence associated with them.

\begin{figure*}[!t]
\includegraphics[width=1.0\textwidth]{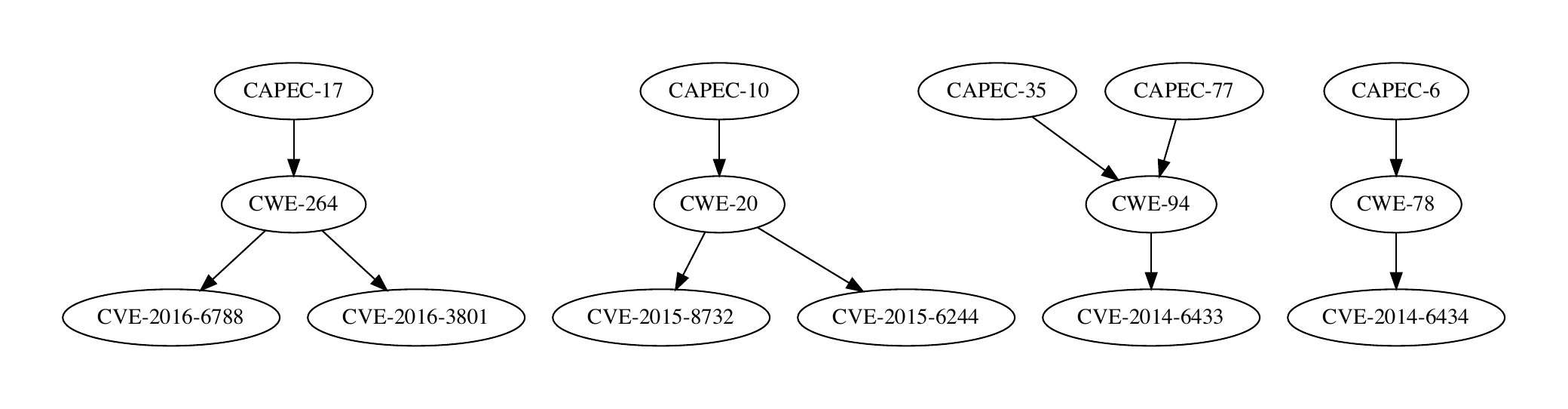}
\caption{An example subset of \(AV\) that only contains relevant evidence, from \(\textit{rel-evidence}\), based on the mission specification \(S\). (Text description of each node is omitted for visualization purposes.)}
\label{fig:AV}
\end{figure*}

For the UAV mission, we produce the graphs \(R\), \(F\), \(\Sigma\), and \(S\) but present only the ones applicable to the vulnerability analysis and impact trace, namely, \(\Sigma\) and \(S\) (figure~\ref{fig:Sigma} and figure~\ref{fig:S}). Using the information present in graph \(AV\) we take the descriptors of \(S\), \(\mathcal{D}_S\) and through \(\textit{evidence}\) and \(\textit{rel-evidence}\) we produce the potential attacks (figure~\ref{fig:AV}). 

We now examine the system structure \(\Sigma\) and mission specification, \(S\). At this stage the analyst can already observe several paths that are going to cause partial or full mission degradation. The final piece of information to find the subsystems most critical to the mission is the subgraph of \(AV\) that's constructed through the stakeholder elicitation and further refined using the system architecture (figure~\ref{fig:AV}). This leads us to segment out three subsystems: (1) Global Positioning System (GPS), (2) XBee radio communication, and (3) GoPro Hero5 camera. The first is concerned with the flight control system. The second is concerned with the ground control station, flight control system, and imagery payload. The third is concerned with just the imagery payload. In a different simulated mission we might have concluded that a different set of subsystems was critical for the same system architecture, \(\Sigma\).

\paragraph{GPS} The attack vectors associated with the GPS are CVE-2016-6788 and CVE-2016-3801. While both of them are targeting the Android operating system it is possible to misconfigure the implementation of the UAV to be exploited in such a manner. Since our analysis starts early and is exercised often in a system's lifecycle we can be conservative with what we consider a vulnerability, so that we can report it back to the stakeholders. The first attack vector can violate the communication based on I2C based on the Mediatek 3339 present in the GPS and the drivers that are necessary for it to operate. The second has to do with crafting an application to increase the attackers privilege level within the system structure. This would require operator action. The set

\begin{equation*}
E_\text{GPS} = \{\emptyset, \text{CVE-2016-6788}, \text{CVE-2016-3801},  \{\text{CVE-2016-6788}, \text{CVE-2016-3801}\},  \text{CWE-264}, \text{CAPEC-17}\}
\end{equation*}

\noindent
contains all relevant evidence from \(\textit{AV}\) for the GPS. It also shows that attacks can be used individually or in combination. Using this set of attacks in combination would yield complete violation of the GPS but also the primary microcontroller ARM STM32F4. Hence the two attacks used in sequence would construct the attack chain,

\[p_\text{GPS} = (\text{GPS} \xrightarrow{\{\text{CVE-2016-6788}, \text{CVE-2016-3801}\}} \text{ARM STM32F4})\].

\noindent
This specific attack chain, if successful, would cause mission degradation by violating the flight control system based on the mission specification \(S\) and would construct the impact trace,

\begin{equation*}
\text{Impact Trace}_{S_\text{GPS}} = ({\text{FCS} \rightarrow \text{CA3.1} \rightarrow \text{SC3.1} \rightarrow \text{H3} \rightarrow \text{L2}}) \sqcup ({\text{FCS} \rightarrow \text{CA3.1} \rightarrow \text{SC3.1} \rightarrow \text{H3} \rightarrow \text{L3}}).
\end{equation*}

\noindent
Furthermore, the CAPEC and CWE entries present in \(E_\text{GPS}\) inform the analyst about classes of weaknesses or general attack patterns that might derive from the reported CVE entries. These can then be encoded more concretely in the requirements documentation and handled at the deployment phase appropriately. In this case, CWE-264 is a category of weaknesses for ``Permissions, Privileges, and Access Controls'' and CAPEC-17 an attack pattern ``Accessing, Modifying or Executing Executable Files.'' This way, the analyst can report the results more concretely to the stakeholders and discuss about potential classes of attack vectors instead of specific instances of them.

\paragraph{XBee} The imagery payload communicates with the ground control station by utilizing the XBee radio module, which in turn requires the use of the ZigBee protocol to communicate to the Beaglebone Black imagery processor. A potential attack that can violate the imagery processor is directly correlated with the drivers it needs to run in order to properly implement the ZigBee IEEE 802.15.4 protocol. Two potential attacks present in relevant evidence from \(AV\) are CVE-2015-8732 and CVE-2015-6244 (figure~\ref{fig:AV}). The two attacks rely on two separate bugs on the driver implementation but have the same causal effect, namely, that an attacker can send a set of packets that cause out-of-bound read and application crashes, which results in a successful denial-of-service. The corresponding CWE-20 ``Improper Input Validation'' and CAPEC-10 ``Buffer Overflow via Environment Variables'' further define the possible type of violation. Similarly with the GPS the evidence is,

\begin{equation*}
E_\text{XBee} = \{\emptyset, \text{CVE-2015-8732}, \text{CVE-2015-6244}, \text{CWE-20}, \text{CAPEC-10}\}.
\end{equation*}

In this instance the attack chain is reliant on the use of multiple XBee devices,

\[p_\text{XBee} = (\text{Imagery XBee} \xrightarrow{a} \text{GCS XBee} \xrightarrow{a} \text{FCS XBee})\]

\noindent
where \(a = \text{CVE-2015-8732} \vee \text{CVE-2015-6244}\) and the impact traces associated with this violation are:

\begin{equation*}
\begin{split}
\text{Impact Trace}_{S_\text{XBee}} = i_\text{GCS XBee} \sqcup i_\text{FCS XBee} \sqcup i_\text{Imagery XBee} %
\end{split}
\end{equation*}

\noindent
where,
\begin{align*}
i_\text{Imagery XBee} = &(\text{Imagery XBee} \rightarrow \text{CA4.3} \rightarrow \text{SC4.3} \rightarrow \text{SC3.1} \rightarrow \text{H3} \rightarrow \text{L2})\\ &\sqcup (\text{Imagery XBee} \rightarrow \text{CA4.3} \rightarrow \text{SC4.3} \rightarrow \text{SC3.1} \rightarrow \text{H3} \rightarrow \text{L3})\\ &\sqcup (\text{Imagery XBee} \rightarrow \text{CA4.3} \rightarrow \text{SC4.3} \rightarrow \text{SC2.1} \rightarrow \text{H1} \rightarrow \text{L1}) \\ &\sqcup (\text{Imagery XBee} \rightarrow \text{CA4.3} \rightarrow \text{SC4.3} \rightarrow \text{SC2.1} \rightarrow \text{H2} \rightarrow \text{L1}),
\end{align*}
\begin{align*}
i_\text{FCS XBee} = &(\text{FCS XBee} \rightarrow \text{CA4.3} \rightarrow \text{SC4.3} \rightarrow \text{SC3.1} \rightarrow \text{H3} \rightarrow \text{L2}) \\ &\sqcup (\text{FCS XBee} \rightarrow \text{CA4.3} \rightarrow \text{SC4.3} \rightarrow \text{SC3.1} \rightarrow \text{H3} \rightarrow \text{L3})\\ &\sqcup (\text{FCS XBee} \rightarrow \text{CA4.3} \rightarrow \text{SC4.3} \rightarrow \text{SC2.1} \rightarrow \text{H1} \rightarrow \text{L1}) \\ &\sqcup (\text{FCS XBee} \rightarrow \text{CA4.3} \rightarrow \text{SC4.3} \rightarrow \text{SC2.1} \rightarrow \text{H2} \rightarrow \text{L1}), \text{ and}
\end{align*}
\begin{align*}
i_\text{GCS XBee} = &(\text{Ground Control Station} \rightarrow \text{CA2.1} \rightarrow \text{SC2.1} \rightarrow \text{H1} \rightarrow \text{L1})\\ & \sqcup (\text{Ground Control Station} \rightarrow \text{CA2.1} \rightarrow \text{SC2.1} \rightarrow \text{H2} \rightarrow \text{L1}).
\end{align*}

\paragraph{GoPro Hero5} The reconnaissance mission depends on the camera. The GoPro Hero series is reported to be susceptible to hijacking variables on start and restart operations that are controlled by the ground control station and are likely to be executed in flight. There are two attacks from the relevant evidence in \(AV\) applicable to this camera, CVE-2014-6433 and CVE-2014-6434. The first allows any arbitrary code to be run during the start command of the camera. The corresponding CWE-94 ``Improper Control of Generation of Code (Code Injection),'' CAPEC-35 ``Leverage Executable Code in Non-Executable Files,'' and CAPEC-77 ``Manipulating User-Controlled Variables'' provide higher-level description of the attack and its side effects.
The second allows any arbitrary code to be run on the camera during the restart command and, hence, can cause an attack chain with the Beaglebone Black or simply stop its operation in the duration of the mission. Similarly with the first it is associated with CWE-78 ``Improper Neutralization of Special Elements used in an OS Command'' and CAPEC-6 ``Argument Injection.'' Therefore, the evidence

\begin{equation*}
E_\text{XBee} = \{\emptyset, \text{CVE-2014-6434}\}, \text{CWE-20}, \text{CAPEC-10}, \text{CVE-2014-6433}, \{\text{CVE-2014-6434}, \text{CVE-2014-6433}\}\}.
\end{equation*}

\noindent
No attack chain can be deduced from the given relevant evidence for this case study. However, there still exists an impact trace,

\begin{align*}
\text{Impact Trace}_{S_\text{GoPro}} = &(\text{Imagery Payload} \rightarrow \text{SC4.2} \rightarrow \text{SC2.1} \rightarrow  \text{H1} \rightarrow \text{L1}) \\ &\sqcup (\text{Imagery Payload} \rightarrow \text{SC4.2} \rightarrow \text{SC2.1} \rightarrow  \text{H2} \rightarrow \text{L1}).
\end{align*}

\paragraph{Mission Impact} From the analysis above the vulnerable paths are:

\[\text{Vulnerable Path}_\Sigma = p_\text{GPS} \sqcup p_\text{XBee}.\]

Finally, we construct the complete impact trace that is going to be reported to the stakeholders for further analysis,

\begin{equation*}
\text{Impact Trace}_S = \text{Impact Trace}_{S_\text{GPS}} \sqcup \text{Impact Trace}_{S_\text{XBee}} \sqcup \text{Impact Trace}_{S_\text{GoPro}}.
\end{equation*}

In the case of further defenses or architectural changes are present the analysis can be repeated to take them into account. In this use-case, the system contains no specific set of defenses. This means that it is vulnerable to the full set of relevant evidence. This is not an unrealistic assumption even for safety-critical systems, since there exist deployed systems that consider cybersecurity as an after-thought. This ends up promoting reactive approaches to security, while Mission Aware is a proactive methodology that is used to design security by design.

\subsection{Discussion}

Mission Aware demonstrates and reduces the number of subsystems requiring inspection by taking intro account requirements for mission success. The scalability of Mission Aware to complex systems is supported by three key factors:
\begin{enumerate}
\item    Its robust hierarchical modeling and graph metamodel.
\item    The evidence associated with the model.
\item    The implications of this evidence on mission outcomes.
\end{enumerate}

\paragraph{Model}  The model is constructed using a standardized systems engineering language, which allows for comprehensive representation of complex systems. This approach offers several advantages. Its versatility enables the capture of all necessary domains within a single framework, while its flexibility allows for easy modifications and updates, such as refining mission specifications based on defined requirements, behaviors, and system architecture. The language's graphical nature facilitates clear communication and understanding of system relationships, providing a visual representation that enhances comprehension. Furthermore, the model can be seamlessly converted to other formats, such as graph representations, without loss of information. This modeling approach provides a robust foundation for analyzing and designing large-scale systems, ensuring consistency across different stages of development and enabling efficient collaboration among stakeholders.

\paragraph{Evidence}  The evidence collected for a given analysis is tailored to the specific mission and the anticipated threats, eliminating the need for an exhaustive analysis of all possible attack vectors. Mission Aware's model-driven approach, while not relying on a realized system, may generate more evidence than necessary due to the comprehensive nature of system descriptors. This leads to the concept of \emph{relevant evidence}---a subset of database entries deemed truly applicable to the system under consideration. This focused approach results in a significantly smaller set of evidence compared to traditional perimeter-based methods, streamlining the analysis process while maintaining effectiveness. By concentrating on mission-critical elements, Mission Aware enables more efficient and targeted security assessments, potentially reducing the workload for analysts without compromising the depth of threat analysis.

\paragraph{Impact} Assessing mission impact typically doesn't require an exhaustive evaluation of the entire system. The Mission Aware approach, grounded in systems theory, allows analysts to identify critical subsystems with higher confidence. This top-down method, starting from high-level mission hazards, efficiently pinpoints potential vulnerabilities. Ideally, Mission Aware consistently produces a set of vertices $|S|$ that is smaller than the total system structure $|\Sigma|$, where $S$ is derived from $\Sigma$. This means the subsystems directly linked to mission degradation form a more manageable subset of the overall system. By focusing on these key elements, Mission Aware enables a more targeted and efficient analysis, reducing the scope of investigation while maintaining a comprehensive understanding of potential mission impacts.

\paragraph{Limitations} The effectiveness of the methodology is subject to several factors. First, its scalability to extremely large, complex system-of-systems could pose computational challenges in graph generation and traversal, although the hierarchical modeling is designed to mitigate this. Second, the analysis produces a snapshot based on a defined mission; evolving missions or dynamic operational contexts require that the underlying models be diligently maintained and updated. Finally, the evidence-based portion of the analysis relies on public vulnerability databases (e.g., CVE). This makes the approach highly effective at identifying risks from ``known knowns,'' but less equipped to handle novel, zero-day attacks or ``unknown unknowns.''

\paragraph{Future work} The limitations highlight promising avenues for future research. To address the challenge of novel threats, we propose extending the methodology to include a third, cause-agnostic analytical path. This new path would involve simulating the impact of component or communication link impairment on mission success, regardless of the cause—be it a novel cyber-attack, random equipment failure, or an environmental effect. By analyzing the system's response to the functional loss of subsystems (individually and in combination), one can identify critical failure points and develop resilience strategies that are not tied to a specific, known threat vector. This cause-agnostic perspective aligns with a broader shift towards resilience engineering and would provide a more robust defense against the unpredictable nature of advanced adversaries, ensuring mission integrity even when specific threats are not yet known.

\section{Conclusion}

In this paper, we have described a methodology, called Mission Aware, that is grounded in systems  theory as well as evidence-based vulnerability assessment of the resulting system architecture. Mission Aware provides strategic, model-based cybersecurity analysis, where cybersecurity is framed according to the mission that the technical system is intended to fulfill, and what potential hazards the mission might face. 
The key innovations of Mission Aware are:

\begin{enumerate}
\item a formal, traceable modeling framework that captures the mission requirements, admissible behaviors, and architectural elements;
\item model-driven identification of attack vectors that are applicable to elements of system; and
\item reduction in the number of system elements that need to be analyzed in order to assure mission success by explicitly focusing the evidence related to mission objectives.
\end{enumerate}

These innovations are achieved by connecting the mission requirements, the admissible functional behaviors, and the system structure that can fulfill the defined service via a guided elicitation process we termed the stakeholders. Both the elicitation process and the resulting outputs are grounded in systems theory and leverage recent work in safety analysis.

All of this information is formally encoded in a mission specification, which uses set-theoretic and graph-theoretic formalisms. The security posture of the overall mission is then directly traced to the security posture of the critical subsystems that relate to potential mission degradation. We assess the security posture of individual subsystems, and their connections, through the use of vulnerability repositories using CAPEC, CWE, and CVE.

As a final observation we note the experience of using a systematic, model-driven process to conduct vulnerability analysis often yields more information than just quantifying the vulnerability aspects of the system. The process itself is an iterative learning experience, allowing circumspection into how a system behaves in response to potential threats and attacks. Therefore, with mission-oriented perspectives we attempt to bring ``what-if'' analysis across the wide range of stakeholders from command level to acquisition to engineering support. The inclusion of this information into review processes and mission activities can enlighten how managers and operators implement defense capabilities from a mission perspective.

\bibliography{manuscript}

\end{document}